\documentclass[review,12pt]{elsarticle}
\usepackage{lineno,hyperref}
\usepackage{graphicx}
\usepackage{geometry}
\usepackage{caption}
\usepackage{subcaption}
\usepackage[linesnumbered,ruled,lined]{algorithm2e}
\usepackage{float}
\usepackage{tikz}
\usepackage{epstopdf}
\usepackage{amssymb}
\usepackage{amsmath}
\usepackage{indentfirst}
\usepackage{subcaption}
\usepackage{footnote}
\usepackage{array}
\usepackage{cases}
\usepackage{multirow}
\usepackage{ upgreek }
\usepackage{makecell}
\begin{document}
%
%
\begin{frontmatter}
\title{An efficient truncation scheme for Eulerian and total Lagrangian SPH methods}

\author[1]{Zhentong Wang}
\ead{zhentong.wang@tum.de}

\author[2]{Chi Zhang}
\ead{zhangchi0118@gmail.com}

\author[1]{Oskar J. Haidn}
\ead{oskar.haidn@tum.de}

\author[1]{Xiangyu Hu\corref{cor1}}
\ead{xiangyu.hu@tum.de}

\address[1]{TUM School of Engineering and Design, Technical University of Munich, Garching, 85747, Germany}

\address[2]{Huawei Technologies Munich Research Center, Munich, Germany}

\cortext[cor1]{Corresponding author: Xiangyu Hu}

\begin{abstract}
In smoothed particle hydrodynamics (SPH) method, 
the particle-based approximations are implemented via kernel functions, 
and the evaluation of performance involves two key criteria: numerical accuracy and computational efficiency.
In the SPH community, 
the Wendland kernel reigns as the prevailing choice due to its commendable accuracy and reasonable computational efficiency. 
Nevertheless, 
there exists an urgent need to enhance the computational efficiency of numerical methods while upholding accuracy.
In this paper, we employ a truncation approach to limit the compact support of the Wendland kernel to 1.6h. 
This decision is based on the observation that particles within the range of 1.6h to 2h make negligible contributions, practically approaching zero, to the SPH approximation.
To address integration errors stemming from the truncation, 
we incorporate the Laguerre-Gauss kernel for particle relaxation due to the fact that this kernel has been demonstrated to enable the attainment of particle distributions with reduced residue and integration errors \cite{wang2023fourth}.
Furthermore, 
we introduce the kernel gradient correction to rectify numerical errors from the SPH approximation of kernel gradient and the truncated compact support.
A comprehensive set of numerical examples including fluid dynamics in Eulerian formulation and solid dynamics in total Lagrangian formulation are tested and have demonstrated that truncated and standard Wendland kernels enable achieve the same level accuracy but the former significantly increase the computational efficiency.
\end{abstract}

\end{frontmatter}
%
%
\section{Introduction}\label{referebce}
Smoothed particle hydrodynamics (SPH) emerges as a meshless methodology, 
initially conceptualized by Lucy \cite{lucy1977numerical}, 
Gingold and Monaghan \cite{monaghan1994simulating} and has found extensive traction across domains, 
including fluid dynamics \cite{monaghan1994simulating}, solid mechanics \cite{libersky1993high}, 
and various scientific and industrial applications \cite{randles1996smoothed, longshaw2015automotive, wang2023eulerian}.
Given that SPH relies on particle-based approximation facilitated by Gaussian-like kernel functions within compact support, 
the choice of the kernel functions significantly affects both the accuracy and computational efficiency.
The accuracy attained within SPH simulations depends on numerical errors which encompass the smoothing error denoted as $E_s$ 
and the integration error represented by $E_r$ \cite{litvinov2015towards}, 
where the smoothing error is introduced in the process of replacing the Dirac delta function with a smooth kernel function 
and the error order is defined by vanishing moments \cite{litvinov2015towards}, 
and the integration error is introduced in the process of discretizing the continuous volume into the summation of the volumes of all neighboring particles. 
Importantly, 
the cumulative truncation error can be expressed as $E_t = E_s + E_r$.
In terms of computational efficiency, 
the size of the cut-off radius determines the number of neighbouring particles, 
and different kernels with different cut-off sizes have a noticeable effect on the computational effort.

Within the SPH community, kernels such as Wendland \cite{wendland1995piecewise}, 
Gauss \cite{gingold1977smoothed}, 
and B-Spline \cite{johnson1996sph} are widely recognized and acknowledged and generally possess the characteristic of satisfying the first vanishing moment \cite{litvinov2015towards}, 
ensuring the second-order smoothing error $E_{s}=\mathcal{O}(h^2)$.
While Gauss kernel as a nature choice for SPH having the advantage of good accuracy \cite{morris1996analysis}, 
the compact support must be truncated to $3h$ during implementation to ensure a sufficiently small integration error 
and the truncated compact support is larger than the non-truncated compact support of other kernels, leading to poor computational efficiency.
Moreover, 
Gauss kernel with truncated compact support, 
similar to B-spline kernel, 
leads to a randomly disordered particle system, 
where the particles exhibit a tendency to cluster together (known as the pair-instability problem).
Thus, in recent years, 
Wendland kernel with the compact support of $2h$ has become the mainstream 
due to its capacity to overcome the pair-instability problem and maintain high computational efficiency 
while achieving the same level of accuracy as the Gauss kernel.
For the given kernel, 
enhancing the computational efficiency in SPH can be achieved by decreasing the resolution, 
i.e. reducing the total number of particles in the whole computational domain, 
at the cost of decreasing the accuracy. 
Therefore, 
the challenge lies in improving computational efficiency without compromising the accuracy.

It is widely recognized that particle distribution has a significant impact on the integration error \cite{quinlan2006truncation,wang2023fourth} 
and obtaining a high-quality particle distribution becomes imperative and critical in preserving numerical accuracy.
In practical applications, 
particle relaxation has emerged as a remarkably effective technique for generating body-fitted particles characterized by complex geometries. 
While the utilization of Wendland kernel for particle relaxation circumvents the pair-instability problem 
and resembles typical liquid molecules in a microscopic description \cite{litvinov2015towards,kirkwood1942radial}, 
the quest for higher-quality particle distribution leads to the introduction of Laguerre-Gauss kernel investigated by Wang \cite{wang2023fourth} as a replacement for Wendland kernel. 
In his work, 
Wang \cite{wang2023fourth} proves that the Laguerre-Gauss kernel has the property of achieving a substantially smoother particle distribution, 
similar to the lattice particle distribution, with significantly reduced relaxation residue in the relaxation process.
Besides, 
by analyzing the profile of Wendland kernel function, 
it is evident that the value of the kernel function rapidly diminishes as the distance from the center particle increases. 
At a distance of $1.6h$, the value almost approaches zero, 
rendering the contribution of all particles within the $1.6h$ to $2h$ range virtually negligible. 
Drawing inspiration from this observation, 
we opt to exclusively utilize particles within the truncated support domain to $1.6h$ for SPH approximation, 
leading to the reduction of the total neighboring particles in this compact support and a significant improvement in computational efficiency.
Note that even though the kernel values of particles within the support domains from $1.6h$ to $2h$ close to zero leading to little effect on the truncation error intuitively, 
it is necessary to quantitatively analyze the impacts of truncating this region on the truncation error. 

In this paper, 
we conducts a quantitative analysis of truncation errors employing both the standard Wendland kernel (with full compact support as $2h$) and the truncated Wendland kernel (with compact support as $1.6h$). 
This analysis is performed with and without the kernel gradient correction across three distinct particle distributions, 
which encompass the lattice distribution and the relaxed distributions utilizing both the standard Wendland and the truncated Laguerre-Gauss kernels for particle relaxation.
Furthermore, 
the truncated Wendland kernel is implemented in a series of numerical examples, 
spanning fluid dynamics within the Eulerian SPH framework and solid dynamics in the total Lagrangian SPH framework to assess the numerical accuracy and computational efficiency in comparison to the standard Wendland kernel.

The structure of paper begins with an in-depth presentation of SPH approximation and the characteristics of kernel in Chapter \ref{Truncated kernel for SPH method}. 
In particular, 
the truncated Wendland kernel is discussed in detail and subjected to a quantitative analysis. 
Moving on to Chapter \ref{Eulerian compressible and weakly compressible SPH}, 
the conservation equation and its corresponding discretization in Eulerian SPH method are provided. 
In addition,  
the Riemann solvers, 
dissipation limiters \cite{toro1994restoration,toro2019hllc,toro2013riemann,wang2023extended} and the kernel gradient correction \cite{zago2021overcoming} are introduced. 
The governing equations and discretization forms in total Lagrangian SPH method are also explained. 
In Chapter \ref{Numerical results}, 
a series of numerical examples are used to meticulously investigate the numerical accuracy and computational efficiency of the truncated Wendland kernel.
%
%
\section{Truncated kernel for SPH method}\label{Truncated kernel for SPH method}
%
%
\subsection{SPH approximations}\label{SPH approximations}
In SPH method, 
the gradient of a function $f(\mathbf{r})$  
at the position $\mathbf{r}_{i}$ can be approximated as 
\begin{equation}
\label{smoothing error}
\begin{split}
\nabla f(\mathbf{r}_{i}) \approx \int \limits_\Omega \nabla f(\mathbf{r}) W(\mathbf{r}_{i}-\mathbf{r}, h)dV = -\int \limits_\Omega f(\mathbf{r}) \nabla W(\mathbf{r}_{i}-\mathbf{r}, h)dV \approx -\sum_{j} f(\mathbf{r}_j) \nabla W_{ij} V_{j}.
\end{split}
\end{equation}
Here, 
$V_j$ is the particle volume and $W_{ij}=W(\mathbf{r}_{i}-\mathbf{r}_{j}, h)$, and the gradient of kernel $\nabla W_{i j}=\frac{\partial W_{i j}}{\partial r_{ij}}\mathbf{e}_{ij}$.
Note that the first approximation of Eq. \eqref{smoothing error} introduces the smoothing error $E_s$ due to the smooth kernel function replacing the Dirac delta function, and the second approximation  introduces the integration error $E_r$, and then the truncation error $E_t$ with $E_t=E_s+E_r$.
Following Ref.\cite{zhang2022smoothed}, 
a weak form of $\nabla f(\mathbf{r}_{i})$ can be rewritten as 
\begin{equation}\label{weak form}
	\nabla f_{i}=\nabla f_{i}+f_{i}\nabla 1 \approx -2\sum_{j} \bar{f}_{ij} \nabla W_{ij} V_{j}
\end{equation}
to discretize the conservation equations due to its conservation property.
Here, 
$\bar{f}_{ij}=(f_{i}+f_{j})/2$ denotes the particle average value of particles $i$ and $j$. Note that Eq. \eqref{weak form} is employed in subsequent sections for the error estimation

%
%
\subsection{Properties of kernel function}
As discussed in Ref. \cite{wendland1995piecewise,liu2010smoothed,monaghan1992smoothed,yang2014new}, a smooth kernel function must satisfy several conditions which are three important properties:

$(1)$ Normalization condition, 
i.e unity condition, requires that the kernel function satisfies 
\begin{equation}
\label{normalization rule}
	\int \limits_V W(\mathbf{r}, h)dV=1. 
\end{equation}

$(2)$ When the smooth length tends to 0, the kernel function follows the Dirac function property, that is 
\begin{equation}
\label{smoothing rule}
	\lim_{h \to 0}W(\mathbf{r}, h) = \delta(\mathbf{r}).
\end{equation}

$(3)$ The compact condition requires
\begin{equation}
\label{Compact condition}
	W(\mathbf{r}, h)=0, \text{when} \left|\mathbf{r}\right| \textgreater \kappa h ,
\end{equation}
where $\kappa$ determine the size of compact support.
%
%
\subsection{Truncated Wendland kernel}
%
%
\subsubsection{Truncated kernel basics}
\begin{figure}
	\centering
	\includegraphics[width=0.6\textwidth]{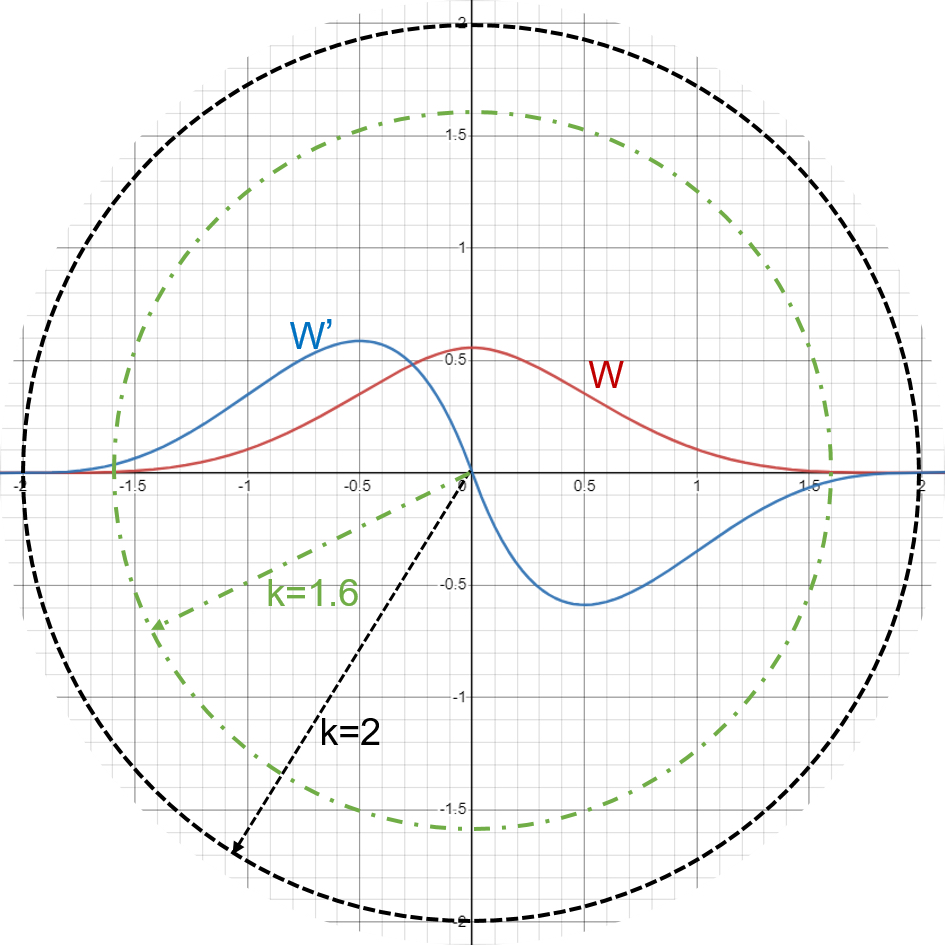}
	\caption{Wendland kernel: function (red) and its derivative (blue) in the compact support and two different compact support size with $\kappa=2$ and $\kappa=1.6$ are plotted as black and green circles, respectively.}
	\label{kernel function and its derivative}
\end{figure}

Following Ref.\cite{wendland1995piecewise}, 
Wendland kernel reads 
\begin{equation}\label{standard kernel}
W(s,h)=\alpha_{d}\begin{cases} (1-s/2)^{4}(2s+1) & \text { if } 0 \leq s \leq \kappa h \\ 0 & \text { if } s>\kappa h \end{cases}, 
\end{equation}
where $s=\left|\mathbf{r}\right|$ and $\alpha_{d}$ is the normalized coefficient in $d$ dimensional space with the value of $\alpha_{1}=3/(4h)$, 
$\alpha_{2}=7/(4\pi h^{2})$ and $\alpha_{3}=21/(16\pi h^{3})$. 
In the SPH community, 
the constant parameter $\kappa$ is usually adopted as $2$, i.e. the cut-off radius is $2h$.
As shown in Figure \ref{kernel function and its derivative}, 
the kernel function curve (red one) and its derivative (blue one) are plotted with the black and green circles representing the compact support domain with cut-off size of $2h$ and $1.6h$, respectively.
It is straightforward to find that $W(s,h)$ is very close to $0$ in the range of support domains where particle distance is larger than $1.5h$, implying particles in this range contribute essentially nothing to the normalization condition. 
Through this observation, we can truncate the compact support size from $2h$ to $1.6h$, which leads to a reduction in the number of neighboring particles, and therefore optimizes computational efficiency.
Note that the Wendland kernel with truncated compact support of $1.6h$ satisfies the normalization condition due to the observation that little distribution of the particles located in the region where particle distance is larger than $1.6h$, which will be discussed quantitatively in section \ref{error analysis}.
While the first derivative of the kernel is not very close to zero in the truncated region, 
the kernel gradient correction matrix therefore introduced in section \ref{preparation} eliminates and compensates the truncation error.
It is worth noting that the present trunction scheme does not affect another two conditions of Eqs. \eqref{smoothing rule} and \eqref{Compact condition}.
%
%
\subsubsection{Preparation of error analysis}
\label{preparation}
In subsequent content, 
the error analysis is undertaken within the prescribed particle distribution,
encompassing both the lattice distribution and the relaxed distributions. 
The particle relaxation technique \cite{zhu2021cad} on the latter distributions is defined by the following expression:
\begin{equation}\label{relaxation acceleration}
    \mathbf{a}_{p,i}=-\frac{2}{\rho}_{i} \sum_{j} p_{0} \nabla W_{i j}V_{j},
\end{equation}
with $p_{0}$ denoting the constant pressure.
It is noteworthy that the acceleration attains the value of zero when the particles attain their steady states.
Besides,
the relaxed distributions in the error analysis contain the particle relaxation using standard Wendland kernel and Laguerre-Gauss kernel \cite{wang2023fourth} which reads
\begin{equation}\label{Laguerre-Gauss kernel expression}
W^{LG}{(s,h)}=\alpha_{d}\begin{cases} (1-\frac{s^{2}}{2}+\frac{s^{4}}{6})e^{-s^{2}} & \text { if } 0 \leq s \leq 2 \\ 0 & \text { if } s>2  \end{cases}, 
\end{equation}
with $\alpha_{d}$ denoting the normalized coefficient where $d$ is dimensional space and the value of $\alpha_{1}=8/(5\times\sqrt{\pi})$, 
$\alpha_{2}=3/\pi$ and $\alpha_{3}=8/\pi^{3/2}$.
Moreover, 
to remedy the truncation error and the approximation of kernel gradient,   
the kernel gradient correction matrix \cite{zago2021overcoming}
\begin{equation}\label{kernel correction matrix}
	\mathbf{B}_i = -\left( \sum_{j} \mathbf r_{ij} \otimes \nabla W_{i j}  V_ {j}\right)^{-1}
\end{equation}
is therefore implemented to improve the numerical accuracy.
%
%
\subsubsection{Error analysis}
\label{error analysis}
To analyze the truncation error, 
we introduce $L_{2}$ error \cite{vergnaud2023investigations} of the particle approximation and its first derivative given by 
\begin{equation}\label{error normalizations}
L_{2}(\phi) = \frac{(\sum \limits_{i} (\left|\phi_{i}^{\text{analytical}}-\phi_{i}^{\text{numerical}} \right|)^2)^{\frac{1}{2}}}{(\sum \limits_{i} (\left|\phi_{i}^{\text{analytical}} \right|)^2)^{\frac{1}{2}}}, 
\end{equation}
where $\phi$ represents the function $f$ or its derivative $df/dx$. 
Here, 
$\phi^{\text{numerical}}$ is approximated as the weak form in Eq.\eqref{weak form}. 
In the case, 
we investigate a circular geometry characterized by a diameter of $D=2$, with a smoothing length of $h=1.3dp$, 
where $dp$ represents the initial particle spacing. 
Three different resolutions, 
namely $dp=D/10$, $dp=D/20$, and $dp=D/40$, 
are employed for the convergence study.
For the clarity throughout our analysis, 
we consider three different particle distributions: 
a lattice distribution, 
a relaxed distribution using the standard Wendland kernel $(\kappa=2.0)$, 
and a relaxed distribution using the Laguerre-Gauss kernel. 
We refer to the errors using the standard Wendland kernel, both with and without the kernel gradient correction, as "SW With B" and "SW Without B", respectively. 
Similarly, 
the errors using the truncated Wendland kernel $(\kappa=1.6)$ are referred to as "TW With B" and "TW Without B", with and without the kernel gradient correction.
 
We firstly explore the normalization condition by investigating the SPH approximation for $\phi$ as a constant function $f(x)=1$ with the resolution $dp=D/10$.
Note that the summation of the kernel weight is independent on the gradient operator, 
and therefore we only consider the "SW without B" and "TW without B".
Table \ref{normalization condition table} lists the errors of SPH approximation for the constant function $f(x)=1$ in different particle distributions with the resolution $dp=D/10$.
The results show the error of the truncated Wendland kernel is almost same as that of the standard kernel, 
validating that the present trunction scheme does not affect the normalization condition of Eqs. \eqref{normalization rule}.
\begin{table}[]
\caption{Error analysis of SPH approximation for an constant function $\phi:f(x)=1$ in different particle distributions with the resolution $dp=D/10$.}
    \centering
\begin{tabular}{ccccc}
\hline
\multirow{2}{*}{Particle distribution}  &  & \multicolumn{3}{c}{$L_{2}(\phi)$}          \\ \cline{3-5} 
                &  & \text{SW without B} &  & \text{TW without B} \\ \hline
\text{Lattice}   &  &   0.010   &  &    0.006  \\
\text{Wendland-based}   &  &   0.034   &  &    0.028  \\
\text{Lagueree-Gauss-based}   &  &   0.037   &  &   0.031  \\
\hline
\end{tabular}
\label{normalization condition table}
\end{table}
As discussed in early studies in Refs. \cite{wendland1995piecewise,litvinov2015towards} that Wendland kernel exhibits 2nd-order smoothing error,  
to investigate the smoothing error of the truncated Wendland kernel, 
we expand function $f(\mathbf{r})$ using Taylor series expansion and rewrite Eq. \eqref{smoothing error} as
\begin{equation}
\label{taylor_series}
\begin{aligned}
&\int \limits_\Omega f(\mathbf{r}) W(\mathbf{r}_{i}-\mathbf{r}, h)dV =\\
&\quad f\left(\mathbf{r}_i\right) \int \limits_\Omega W\left(\mathbf{r}-\mathbf{r}_i, h\right) dV + \nabla f\left(\mathbf{r}_i\right) \int \limits_\Omega \left(\mathbf{r}-\mathbf{r}_i\right) W\left(\mathbf{r}-\mathbf{r}_i, h\right) dV  + \mathcal{O}(h^2).
\end{aligned}
\end{equation}
With the normalization condition and even function condition \cite{yang2014new} in hand,
it is easy to derive 
\begin{equation}\label{smoothing error order}
\begin{aligned}
\int \limits_\Omega f(\mathbf{r}) W(\mathbf{r}_{i}-\mathbf{r}, h)dV 
&=f\left(\mathbf{r}_i\right) + \mathcal{O}(h^2),
\end{aligned}
\end{equation}
indicating that the truncated Wendland kernel remains 2nd-order smoothing error. 
\begin{figure}
	\centering
	\begin{subfigure}[b]{0.8\textwidth}
		\centering
		\includegraphics[trim = 0cm 0cm 0cm 0cm, clip, width=1.0\textwidth]{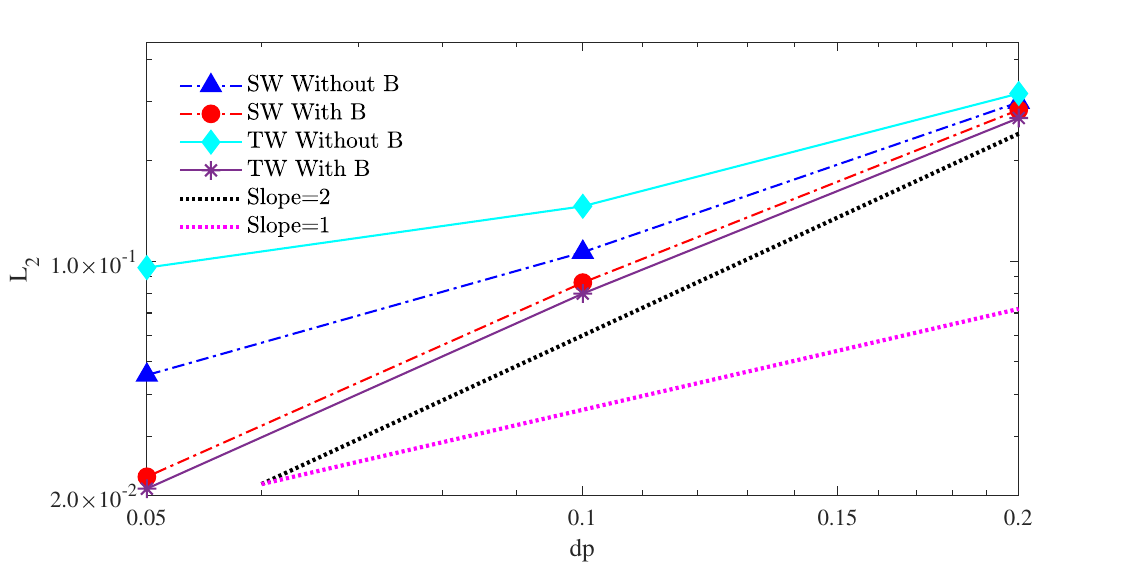}
	\end{subfigure}
	\begin{subfigure}[b]{0.8\textwidth}
		\centering
		\includegraphics[trim = 0cm 0cm 0cm 0cm, clip, width=1.0\textwidth]{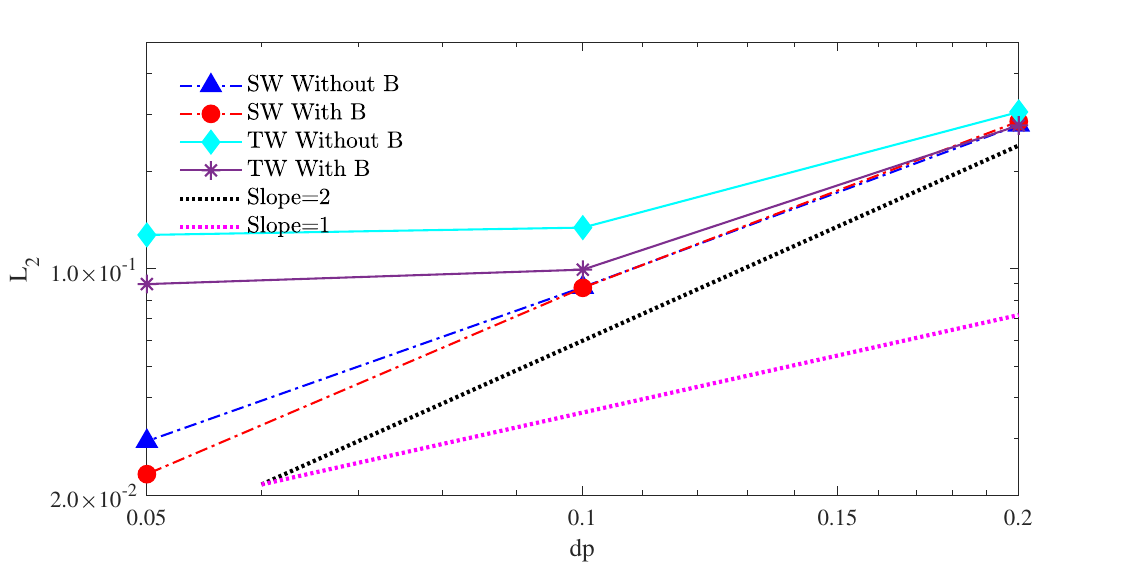}
	\end{subfigure}
 \begin{subfigure}[b]{0.8\textwidth}
		\centering
		\includegraphics[trim = 0cm 0cm 0cm 0cm, clip, width=1.0\textwidth]{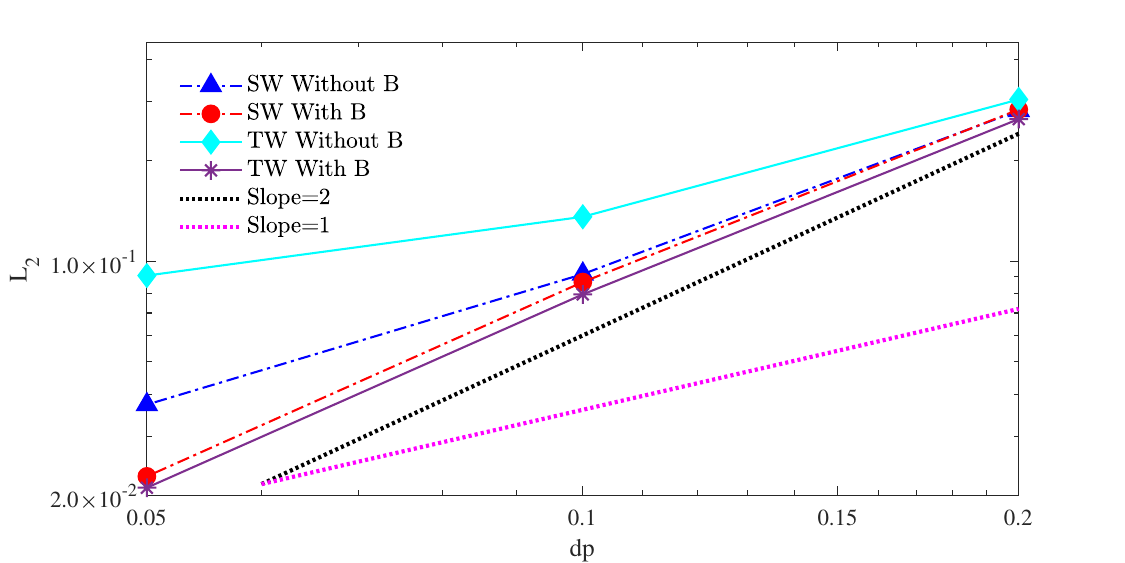}
	\end{subfigure}
	\caption{The convergence study of error for kernel gradient on lattice distribution (upper panel) and the relaxed distributions using Wendland (middle panel) and Laguerre-Gauss (bottom panel) kernels, respectively.}
	\label{convergence study of three particle distributions}
\end{figure}

Moreover, 
in the context of assessing convergence and comparing errors between employing the standard Wendland kernel and the truncated Wendland kernel,
we test the SPH approximation for the derivative of the function $df/dx$ with the exponential function given by
\begin{equation}\label{exp function}
f(x)=e^{-\frac{x^2}{0.1}},
\end{equation}
with the resolutions including $dp=D/10$,$D/20$ and $D/40$.
Figure \ref{convergence study of three particle distributions} illustrates the $L_{2}$ norm of the SPH approximation for the derivative of an exponential function across three distinct particle distributions, 
including the lattice distribution and relaxed distributions employing the Wendland and Laguerre-Gauss kernels and the two lines with scopes equal to $1$ and $2$.

In the top panel of Figure \ref{convergence study of three particle distributions}, 
which pertains to the lattice distribution, it becomes evident that the truncation scheme introduces the extra error by comparing the lines labeled "SW Without B" and "TW Without B".
However, 
the introduction of the kernel gradient correction effectively compensates for the errors originating from SPH approximation and the truncated cut-off radius observed in the lines of "SW With B" and "TW With B". 
Consequently, 
the results using the standard and truncated Wendland kernels with the kernel gradient correction obtain the very similar error and approach nearly second-order convergence.

However, 
in the middle panel of Figure \ref{convergence study of three particle distributions}, 
it is evident that, 
while the kernel gradient correction still reduces errors, 
a clear trend emerges which is that, as the resolution transitions from $dp=D/20$ to $dp=D/40$, 
the error using the truncated Wendland kernel cannot be reduced further. 
It is noteworthy that, 
within the given relaxed distribution using the standard Wendland kernel,
employing the truncation method results in significantly higher errors. 
This is primarily attributed to the truncation scheme's strong reliance on the quality of the particle distribution, specifically, a lower relaxation residue. 
Consequently, 
the relaxed distribution using the standard Wendland kernel is unsuitable for the proposed truncation scheme.

In the bottom panel of Figure \ref{convergence study of three particle distributions}, 
the effect of the kernel gradient correction on error reduction is evident within the given relaxed distribution using the Laguerre-Gauss kernel. 
While the error when employing the truncated Wendland kernel (as seen in the line "TW Without B") is larger than that when using the standard Wendland kernel (as indicated in the line "SW Without B"), 
the kernel gradient correction introduced brings the errors for both the standard and truncated Wendland kernels to the same level and achieves the second-order convergence (as depicted in the lines "SW With B" and "TW With B"). 
This result notably differs from the findings in the middle panel, 
signifying that the particle distribution employing the Laguerre-Gauss kernel is best suited for the truncation scheme.
%
%
\section{Governing equations and SPH discretizations}\label{Eulerian compressible and weakly compressible SPH}
%
%
\subsection{Eulerian SPH for fluid dynamics}\label{Standard Eulerian SPH framework}
%
%
\subsubsection{Governing equations}\label{Eulerian SPH governing equatioins}
The Euler equation can be written as 
\begin{equation}\label{eqs:conservation}
\frac{\partial \mathbf{U}}{\partial t}+\nabla \cdot \mathbf F(\mathbf{U})=0,
\end{equation}
where the conserved variables $\mathbf{U}$ and the corresponding fluxes $\mathbf{F}(\mathbf{U})$ are described as 
\begin{equation}
\label{eqs:flux}
\mathbf{U}=\left[\begin{array}{c}
\rho \\
\rho u \\
\rho v \\
\rho w \\
E
\end{array}\right]\quad \text{and} \quad \mathbf{F}=\left[\begin{array}{c}
\rho u \\
\rho u^{2}+p \\
\rho u v \\
\rho u w \\
u(E+p)
\end{array}\right]+\left[\begin{array}{c}
\rho v \\
\rho vu \\
\rho v^{2}+p \\
\rho vw \\
v(E+p)
\end{array}\right]+\left[\begin{array}{c}
\rho w \\
\rho wu \\
\rho wv \\
\rho w^{2}+p \\
w(E+p)
\end{array}\right],
\end{equation}
where $u$, $v$ and $w$ are the components of velocity, $\rho$ and $p$ are the density and pressure, respectively, and $E=\frac{\rho{\mathbf{v}}^2}{2}+\rho e$ the total energy with $e$ denoting the internal energy.
The equation is closed with an equation of state (EOS).
we apply the EOS of
\begin{equation}\label{getpressure}
    p=\rho(\gamma-1)e,
\end{equation}
with the heat capacity ratio $\gamma=1.4$ for compressible flows whose sound speed is calculated by
\begin{equation}\label{sound equation}
    c^2=\frac{\gamma p}{\rho}.
\end{equation}
For incompressible flow, we apply the artificial EOS
\begin{equation}
p=c^2(\rho-\rho_0). 
\end{equation}
Here, following the weakly-compressible assumption, $\rho_0$ is the reference density and the artificial sound speed $c=10U_{max}$ with $U_{max}$ denoting the maximum flow speed to limit the density variation less than 1\%.
Note that the energy equation described in Eq. \eqref{eqs:flux} is inactive during weakly-compressible flow simulations.
%
%
\subsubsection{Riemann-based Eulerian SPH discretization}\label{Eulerian SPH discretization}
Following Ref. \cite{vila1999particle}, 
the Euler equation can be discretized as 
\begin{equation}\label{eqs:conservation-discretize}
\left\{\begin{array}{l}
\frac{\partial}{\partial t}\left(w_{i}\rho_{i}\right)+2 w_{i}\sum_{j} w_{j}  (\rho \mathbf{v})^{*}_{E, i j} \cdot \nabla W_{i j}=0 \\
\frac{\partial}{\partial t}\left(w_{i}\rho_{i} \mathbf{v}_{i}\right)+
2 w_{i}\sum_{j} w_{j} \left[(\rho \mathbf{v} \otimes \mathbf{v})^{*}_{E, i j}+p^{*}_{E, i j}\mathbb{I}\right] \cdot \nabla W_{i j}=0 \\
\frac{\partial}{\partial t}\left(w_{i}E_{i}\right)+ 2 w_{i}\sum_{j} w_{j} \left[(E\mathbf{v})^{*}_{E, i j}+
(p \mathbf{v})^{*}_{E, i j}\right] \cdot \nabla W_{i j}=0
\end{array},\right.
\end{equation}
where $w$, $\mathbf{v}$ and $\mathbb{I}$ are the volume of particle, velocity and identity matrix, respectively, and terms $()^{*}_{E, i j}$ are solution of the Riemann problem.

For the Riemann solution \cite{toro1994restoration,toro2019hllc,toro2013riemann}, 
the smallest and the largest wave speeds are $S_{l}$ and $S_{r}$, respectively, and the middle speed $S_{\ast}$ is the discontinuity separating the solution into $(\rho_l^{\ast}, u_l^{\ast},p_l^{\ast})$ and $(\rho_r^{\ast}, u_r^{\ast},p_r^{\ast})$.
For compressible flows, we apply the HLLC Riemann solver with $S_{l}$ and $S_{r}$ given by  
\begin{equation}\label{two wave estimation}
\left\{\begin{array}{l}
S_{l}=u_{l}-c_{l} \\
S_{r}=u_{r}+c_{r}
\end{array},\right.
\end{equation}
the middle wave speed 
\begin{equation}\label{middle wave estimation}
S_{\ast}=\frac{\rho_{r} u_{r}\left(S_{r}-u_{r}\right)+\rho_{l} u_{l}\left(u_{l}-S_{l}\right)+p_{l}-p_{r}}{\rho_{r}\left(S_{r}-u_{r}\right)+
\rho_{l}\left(u_{l}-S_{l}\right)}.
\end{equation}
Then, the conserved variables in the intermediate region can be obtained as  
\begin{equation}\label{conserved variables in HLLC}
\left\{\begin{array}{l}
p^{*}=p_{l}+\rho_{l}\left(u_{l}-S_{l}\right)\left(u_{l}-u^{*}\right)=p_{r}+\rho_{r}\left(S_{r}-u_{r}\right)\left(u^{*}-u_{r}\right) \\
\mathbf{v}_{l/r}^{*}=u^{*}\mathbf{e}_{ij}+ \left[\frac{1}{2}(\mathbf{v}_{l}+\mathbf{v}_{r})- \frac{1}{2}({u}_{l}+{u}_{r})\mathbf{e}_{ij}\right]\\
\rho_{l/r}^{*}=\rho_{l/r} \frac{\left(S_{l/r}-\left|\mathbf{v}_{l/r}\right|\right)}{\left(S_{l/r}-u^{*}\right)}\\
E_{l/r}^{*}=\frac{\left(S_{l/r}-\left|\mathbf{v}_{l/r}\right|\right) E_{l/r}-p_{l/r} \left|\mathbf{v}_{l/r}\right|+p^{*} u^{*}}{S_{l/r}-u^{*}}
\end{array},\right.
\end{equation}
with ${u}^{*}={S}_{\ast}$.
For incompressible flows, we apply the linearised Riemann solver \cite{toro2013riemann} with 
\begin{equation}\label{linearised Riemann solver}
\left\{\begin{array}{l}
u^{*}=\frac{u_{l}+u_{r}}{2}+\frac{1}{2} \frac{\left(p_{l}-p_{r}\right)}{\bar{\rho} \bar{c}} \\
p^{*}=\frac{p_{l}+p_{r}}{2}+\frac{1}{2} \bar{\rho}  \bar{c} \left(u_{l}-u_{r}\right)
\end{array},\right.
\end{equation}
where $\bar{\rho}=\frac{1}{2}(\rho_{l}+\rho_{r})$ and $\bar{c}=\frac{1}{2}(c_{l}+c_{r})$

To decrease the numerical dissipation, following Refs. \cite{wang2023eulerian,wang2023extended}, 
the HLLC Riemann solver of Eq. \eqref{conserved variables in HLLC} adopts the dissipation limiter as 
\begin{equation}
\left\{\begin{array}{l}
u^{*}=\frac{\rho_{l}u_{l}c_{l}+\rho_{r}u_{r}c_{r}}{\rho_{l}c_{l}+\rho_{r}c_{r}}+\frac{p_{l}-p_{r}}{\rho_{l}c_{l}+\rho_{r}c_{r}}\beta^{2}_{HLLC}\\
p^{*}=\frac{p_{l}+p_{r}}{2} +\frac{1}{2}\beta_{HLLC}\left[\rho_{r}c_{r}\left(u^{*}-u_{r}\right)-\rho_{l}c_{l}\left(u_{l}-u^{*}\right)\right] 
\end{array}.\right.
\end{equation}
Here, the limiter $\beta_{HLLC}$ reads
\begin{equation}\label{HLLC limiter}
\beta_{HLLC}=\min \left(\upeta_{HLLC} \max (\frac{u_{l}-u_{r}}{\bar{c}}, 0) , 1\right),
\end{equation}
with $\upeta_{HLLC}=1$ according to experimental and numerical examples.
For the linearised Riemann solver of Eq. \eqref{linearised Riemann solver}, the dissipation limiter is introduced as  
\begin{equation}\label{acoustic Riemann solver}
\left\{\begin{array}{l}
u^{*}=\frac{u_{l}+u_{r}}{2}+\frac{1}{2} \beta^{2}_{linearisd} \frac{\left(p_{L}-p_{R}\right)}{\bar{\rho} \bar{c}} \\
p^{*}=\frac{p_{l}+p_{r}}{2}+\frac{1}{2} \beta_{linearisd} \bar{\rho}  \bar{c} \left(u_{l}-u_{r}\right)
\end{array}.\right.
\end{equation}
where the limiter $\beta_{linearisd}$ is defined as 
\begin{equation}\label{acoustic limiter}
\beta_{linearisd}=\min \left(\upeta_{linearisd} \max (\frac{u_{l}-u_{r}}{\bar{c}}, 0), 1\right),
\end{equation}
with $\upeta_{linearisd}=15$.

To compensate the errors introduced from SPH approximation for the kernel gradient and truncated compact support in this paper, 
the kernel gradient correction in Eq. \eqref{kernel correction matrix} is introduced to rewrite the kernel gradient as 
\begin{equation}
	{\nabla}^{'} W_{i j}=\frac{\mathbf{B}_i+\mathbf{B}_j}{2} \nabla W_{i j}
\end{equation}
to replace original $\nabla W_{i j}$ of Eq. \eqref{eqs:conservation-discretize}.
%
%
\subsection{Total Lagrangian SPH for solid dynamics}\label{Standard total Lagrangian SPH framework}
For solid dynamics, the conservation equations in total Lagrangian framework read 
\begin{equation}\label{total Lagrangian SPH governing equatioins}
\left\{\begin{array}{l}
\rho=J^{-1} \rho^0 \\
\rho^0 \frac{d\boldsymbol{v}_{i}}{dt}=\nabla^0 \cdot \mathbb{P}^{\mathrm{T}}
\end{array},\right.
\end{equation}
with $\rho$, $\mathbb{P}$ and $T$ are the density, the first Piola-Kirchhoff stress tensor and the matrix transposition operator, respectively. 
The superscript $0$ denote the value in underformed reference.
Also, 
$J=det(\mathbb{F})$ with $\mathbb{F}$ being the deformation gradient tensor. 
Then, 
The first Piola-Kirchhoff stress tensor $\mathbb{P}=\mathbb{F}\mathbb{S}$ with $\mathbb{S}$ denoting the second Piola-Kirchhoff stress tensor given by \cite{zhang2021sphinxsys}
\begin{equation}
\begin{aligned}
\mathbb{S} =\lambda \operatorname{tr}(\mathbb{E}) \mathbb{I}+2 G \mathbb{E}.
\end{aligned}
\end{equation}
Here,  $\lambda$ and $G$ the Lam$\Grave{e}$ parameter and the shear modulus, respectively.

Following Ref. \cite{zhang2021sphinxsys}, 
the momentum equation of Eq. \eqref{total Lagrangian SPH governing equatioins} can be discretized in total Lagrangian SPH form as 
\begin{equation}\label{total Lagrangian SPH governing equatioins discretization}
\rho_i^0 \frac{d\boldsymbol{v}_{i}}{dt}=\sum_j\left(\mathbb{P}_i \mathbf{B}_i^{0^{\mathrm{T}}}+\mathbb{P}_j \mathbf{B}_j^{0^{\mathrm{T}}}\right) 
\cdot \nabla^0 W_{i j} V_j^0,
\end{equation}
where $\nabla^0 W_{i j}=\frac{\partial W\left(\boldsymbol{r}_{i j}^0, h\right)}{\partial {r}_{i j}^0} \boldsymbol{e}_{i j}^0$ 
and $\mathbf{B}^{0}$ present the kernel gradient and kernel correction matrix in the initial configuration, respectively. 
It is worth noting that the deformation tensor can be updated by using its change rate defined as  
\begin{equation}\label{total Lagrangian SPH governing equatioins discretization}
\frac{d\mathbb{F}}{dt}=\sum_j V_j^0 \left(\boldsymbol{v}_{j}-\boldsymbol{v}_{i}\right) \cdot \nabla^0 W_{i j} \mathbf{B}_i^{0}.
\end{equation}
%
%
%
\section{Numerical results}\label{Numerical results}
In this section, 
we test a set of numerical cases including fluid and solid dynamics to verify the numerical accuracy and computational efficiency of the proposed truncated Wendland kernel. 
Note that the smoothing length is applied as $h=1.3dp$ in fluid dynamics and $1.15dp$ in solid dynamics.
For clarity, 
the Eulerian and total Lagrangian methods using standard Wendland kernel with cut-off radius as $2h$ (denoted as SW) and truncated Wendland kernel with cut-off radius as $1.6h$ (denoted as TW), respectively.
In all cases, 
the computations are all performed on an Intel Core i7-10700 2.90 GHz 8-core desktop computer and the CPU wall-clock times required by standard and truncated Wendland kernel denoted as $\text{T}_\text{SW}$ and $\text{T}_\text{TW}$, respectively.
Here, 
we establish the parameter of computational time rate, denoted as $\alpha$, which is given by the ratio of the CPU wall-clock times using TW to those using SW, i.e. $\alpha = {(\text{CPU wall-clock times})_\text{TW}}/{(\text{CPU wall-clock times})_\text{SW}}$. A lower value of this parameter indicates enhanced computational efficiency when utilizing the TW.
%
%
\subsection{Double Mach reflection of a strong shock}
In this section, 
we undertake a two-dimensional investigation, 
specifically the double Mach reflection phenomenon resulting from a high-strength shockwave. 
This examination serves the purpose of a correctness analysis by using SW and TW in the context of compressible flow dynamics.
\begin{figure}
	\centering
	\begin{subfigure}[b]{0.8\textwidth}
		\centering
		\includegraphics[trim = 0cm 0cm 0cm 0cm, clip, width=1.0\textwidth]{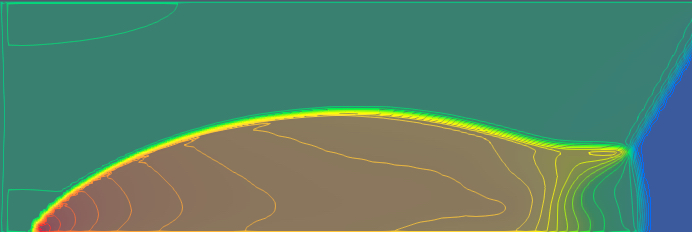}
	\end{subfigure}
	\begin{subfigure}[b]{0.8\textwidth}
		\centering
		\includegraphics[trim = 0cm 0cm 0cm 0cm, clip, width=1.0\textwidth]{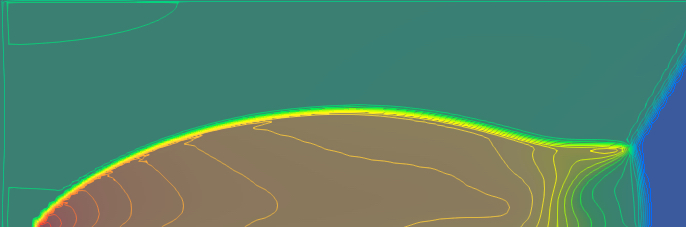}
	\end{subfigure}
	\caption{Double Mach reflection of a strong shock: 
	Density contour and its zoom-in view ranging from $1.3$ to $24.0$ obtained by SW (top panel) and TW (bottom panel)
	with the resolution $dp=1/125$.}
	\label{Density contour ranging from 0 to 25.0 obtained by sw and tw.}
\end{figure}
Following Ref. \cite{woodward1984numerical}, 
we define our computational domain as $(x, y) \in [0, 4] \times [0, 1]$ and the initial conditions are prescribed as follows
\begin{equation}
	(\rho,u,v,p)= \begin{cases}(1.4,0,0,1) &  y\leq1.732(x-0.1667) \\ (8,7.145,-4.125,116.8333) & \text { otherwise }\end{cases}.
\end{equation}
The simulation is performed until the final time $t = 0.2$.
Furthermore, there exists an initial configuration wherein a Mach 10 shockwave propagates to the right, its starting point being $(x, y) = (0.1667, 0)$, and it maintains a 60-degree inclination relative to the x-axis.
The computational domain is bounded by distinct conditions:
$(1)$ The bottom boundary, spanning from $x = 1/6$ to $x = 4$, is defined as a reflective wall boundary;
$(2)$ The left-hand boundary is governed by the post-shock boundary condition;
$(3)$ The right boundary at $x = 4$ adheres to a zero-gradient condition.
Besides, 
the spatial resolution is applied as $dp = 1/125$.

Figure \ref{Density contour ranging from 0 to 25.0 obtained by sw and tw.} shows the density contour ranging from $1.3$ to $24.0$ 
obtained by SW and TW with the resolution $dp=1/125$ at the finial time $t=0.2$, 
indicating that both methods successfully capture key flow features, such as the Mach stem and the near-wall jet and obtain equivalent accuracy of the results.
Moreover, 
the CPU wall-clock times demanded by SW and TW throughout the entire process with the resolution $dp=1/125$ are $226.20s$ and $137.27s$, respectively. 
This indicates that the computational time using the latter is approximately 60\% that of the former, highlighting the significantly improved computational efficiency of the latter over the former.
%
%
\subsection{Double shear layer flow}
In this section, 
the simulation involves a double periodic shear layer in an inviscid flow $(\mu = 0)$ across a square domain with unit edge length. The initial flow depicts a shear layer with finite thickness and a minor vertical velocity perturbation and the initial velocity field \cite{bell1989second,wang2023eulerian} is defined as follows:
\begin{equation}
    u_x = \begin{cases}
    \tanh((y-0.25) / \rho), & \text{for } y \leq 0.5, \\
    \tanh((0.75-y) / \rho), & \text{for } y > 0.5,
    \end{cases}
    \quad \text{and} \quad
    u_y = \delta \sin(2 \pi x),
    \end{equation}
where $\rho=1/30$ and $\delta =0.05$.
\begin{figure}
    \centering
    \includegraphics[width=1.0\textwidth]{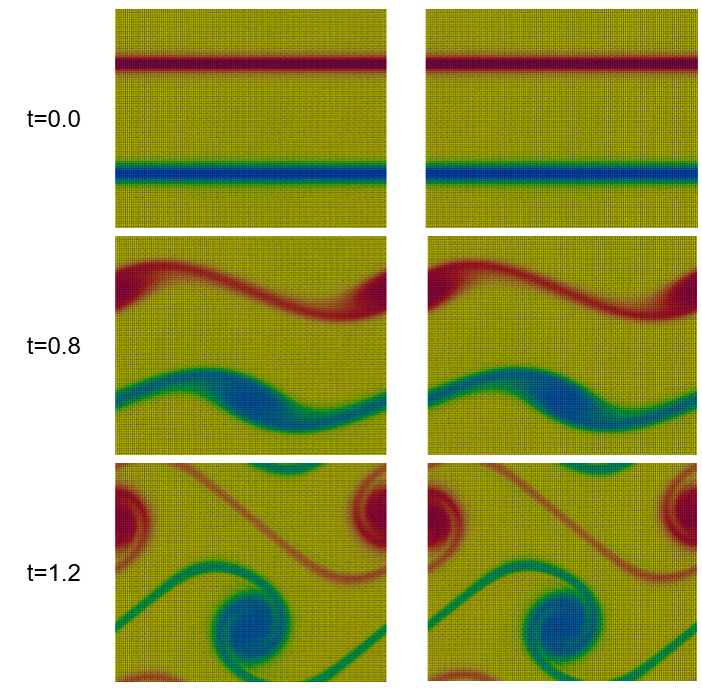}
    \caption{Double shear layer flow : The vorticity contour ranging from $-25$ to $25$ obtained by SW (left panel) and TW (right panel) with the spatial resolution $dp=1/128$ at different times $t=0.0$, $0.8$ and $1.2$.}
    \label{Double shear layer flow vorticity contour}
\end{figure}

Figure \ref{Double shear layer flow vorticity contour} presents the vorticity contour ranging from $-25$ to $25$ obtained by SW and TW with $dp=1/128$ at different times $t=0.0$, $0.8$ and $1.2$.
It can be noted that the truncated kernel is able to obtain results as smooth as the standard kernel, and the results are highly similar, 
which proves that kernel with or without truncation can obtain the same accuracy.
Besides, 
the CPU wall-clock times required by SW and TW in the whole process under the resolution $dp=1/128$ are $29.99s$ and $17.83s$, respectively, and then $\alpha$ is calculated as $0.59$, 
illustrating that the considerably improved computational efficiency of using the proposed kernel.
%
%
\subsection{Lid-driven cavity flows with two shapes}
In this section, 
we delve into two-dimensional lid-driven cavity flows with varying geometries to conduct a computational efficiency comparison between SW and TW. 
Initially, we employ a straightforward square cavity, with the computed results subjected to scrutiny against the benchmark results presented by Ghia et al. \cite{ghia1982high} to validate their accuracy.
Furthermore, we assess the capability of the proposed kernel in handling intricate geometries by conducting simulations on a semi-circular cavity. The outcomes of this simulation are compared with the reference results obtained by Glowinski et al. \cite{glowinski2006wall}. Here, in both situations, the simulations are run until the final time $t=30$ and the upper moving wall is assigned a constant velocity $U_{wall}=1.0$, while all other boundaries are maintained as non-slip wall conditions.
%
%
\subsubsection{Lid-driven square cavity flow}
\begin{figure}
    \centering
    \includegraphics[width=1.0\textwidth]{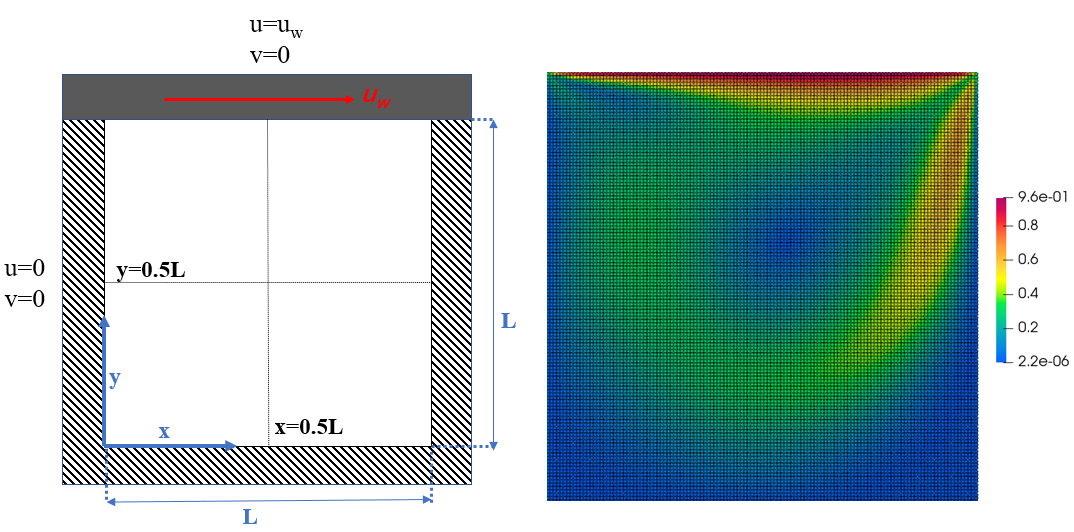}
    \caption{Lid-driven square cavity flow: The square geometry and boundary conditions. Besides, the velocity contour ranging from $2.2 \times 10^{-5}$ to $0.96$ obtained by TW with the resolution as $dp=1/1/129$ under the Reynolds number $Re=400$.}
    \label{Lid-driven square cavity flow square Geometry}
\end{figure}

For the square cavity, 
the geometry and boundary condtions are shown in Figure \ref{Lid-driven square cavity flow square Geometry} and three different spatial resolutions $dp=1/33$, $1/65$ and $1/129$ are adopted for the convergence study.
Also, 
the Reynolds number $Re=400$ is applied in this case.
From the Figure \ref{Lid-driven square cavity flow square Geometry} (right panel), 
it can be noted that the proposed kernel enables obtain the smooth results which are in good agreement with the reference \cite{ghia1982high}.
\begin{figure}
    \centering
    \includegraphics[width=1.0\textwidth]{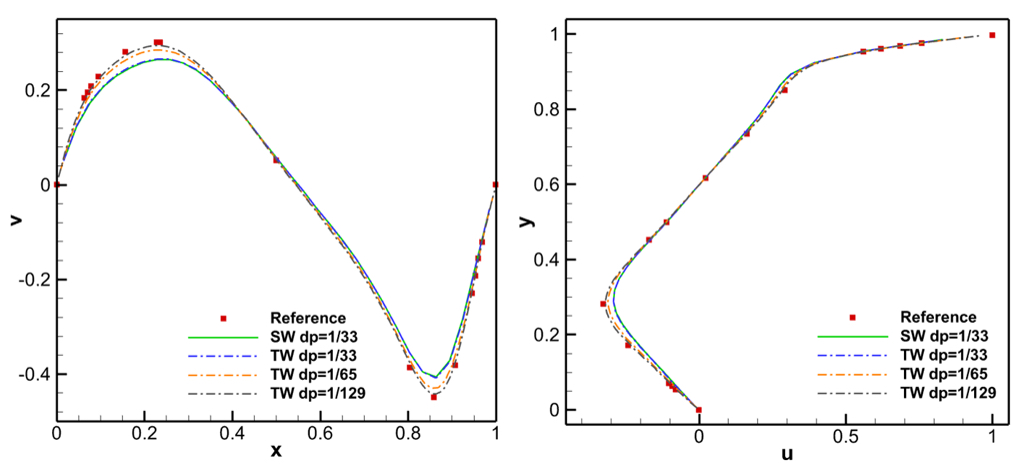}
    \caption{Lid-driven square cavity flow ($Re=400$): The horizontal velocity component $u$ along $x = 0.5L$ (left panel) and 
	the vertical velocity component $v$ along $y = 0.5 L$ (right panel) using SW with the resolution of $dp=1/33$ and TW with the resolutions of $dp=1/33$, $1/65$ and $1/129$, 
	and the comparisons with the reference by Ghia \cite{ghia1982high}.}
    \label{Lid-driven square cavity flow convergence}
\end{figure}
\begin{table}[]
\caption{Lid-driven square cavity flow ($Re=400$): Comparison of computational efficiency using SW and TW with different spatial resolutions.}
 \centering
\begin{tabular}{cccc}
\hline
Resolution     & Kernel type & CPU wall-clock time (s) & $\alpha$   \\ \hline
\multirow{2}{*}{1/33}  & SW      &    6.36  & \multirow{2}{*}{0.76} \\
                       & TW      &    4.81   &         \\
\multirow{2}{*}{1/65}  & SW      &    44.12  & \multirow{2}{*}{0.68} \\
                       & TW      &    30.00  &                      \\
\multirow{2}{*}{1/129} & SW      &    371.78  & \multirow{2}{*}{0.60} \\
                       & TW      &    222.71   &     \\ \hline
\end{tabular}
\label{Lid-driven square cavity flow:computational efficiency.}
\end{table}
Besides, 
Figure \ref{Lid-driven square cavity flow convergence} presents the horizontal velocity component $u$ along $x = 0.5L$ and the vertical velocity component $v$ along $y = 0.5 L$ using SW and TW with the different resolutions, and the comparisons with the reference by Ghia \cite{ghia1982high}.
It can be observed that employing TW allows to achieve the same level of accuracy as SW under the resolution of $dp=1/33$,
indicating that the truncation method has a negligible impact on numerical accuracy.
Also, 
the results of using the proposed kernel achieve the second-order convergence with the increase of the spatial resolutions.
Furthermore, 
the CPU wall-clock times using SW and TW required in the whole process are listed in Table \ref{Lid-driven square cavity flow:computational efficiency.}, 
clearly verifying that the truncation method has significant improvement in the computational efficiency.
%
%
\subsubsection{Lid-driven semi-circular cavity problem}
In this section, we examine more intricate semi-circular geometries with a diameter of $1$ to validate the efficacy of the proposed kernel in obtaining precise and smooth results in flows characterized by complex geometries and the resolution is applied as $dp=1/129$ to discretize the computational domain as well as the Reynolds number $Re=1000$ is adopted.
\begin{figure}
    \centering
    \includegraphics[width=1.0\textwidth]{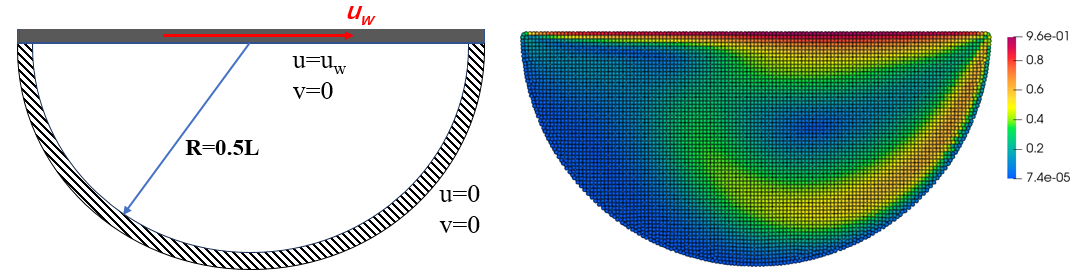}
    \caption{Lid-driven semi-circular cavity flow: The semi-circular geometry and boundary conditions. Besides, the velocity contour ranging from $7.4 \times 10^{-5}$ to $0.96$ obtained by TW with the resolution as $dp=1/1/129$ under the Reynolds number $Re=1000$.}
    \label{Lid-driven semi-circular cavity flow geometry}
\end{figure}
\begin{figure}
    \centering
    \includegraphics[width=1.0\textwidth]{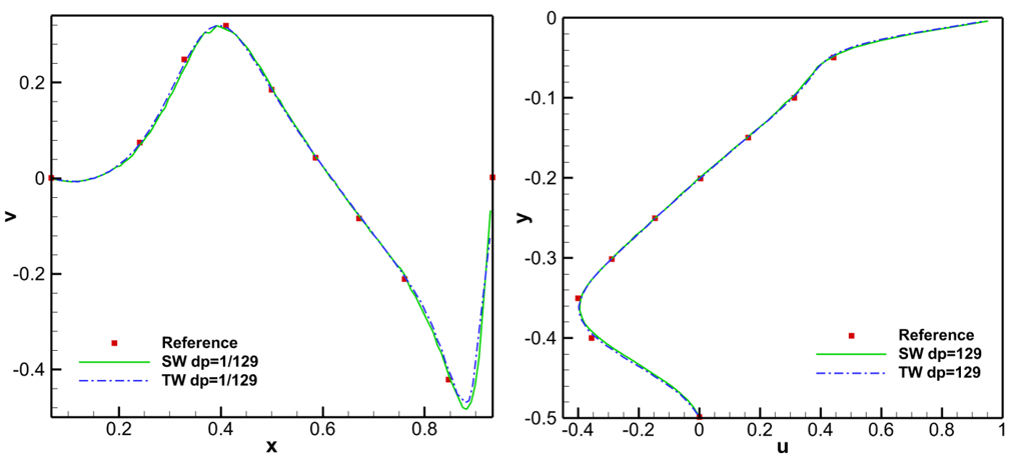}
    \caption{Lid-driven semi-circular cavity flow ($Re=1000$): The horizontal velocity component $u$ along $x = 0.5L$ (left panel) and 
	the vertical velocity component $v$ along $y = -0.25 L$ (right panel) using SW and TW with the resolution of $dp=1/129$, 
	and the comparisons with the reference by Glowinski et al. \cite{glowinski2006wall}.}
    \label{Lid-driven semi-circular cavity flow convergence}
\end{figure}

Figure \ref{Lid-driven semi-circular cavity flow geometry} presents the semi-circular geometry with the radius as $0.5$ and boundary conditions and the velocity contour ranging from $7.4 \times 10^{-5}$ to $0.96$ using TW with the resolution as $dp=1/1/129$ under the Reynolds number $Re=1000$, 
showing that the velocity contour obtained by TW is smooth.
Also, 
Figure \ref{Lid-driven semi-circular cavity flow convergence} illustrates the horizontal velocity component $u$ along $x = 0.5L$ and 
the vertical velocity component $v$ along $y = -0.25 L$ using SW and TW with the resolution as $dp=1/129$, 
and the comparisons with the reference by Glowinski et al. \cite{glowinski2006wall}, 
validating that truncated or standard kernels can obtain the same level of the accuracy in the high resolution.
Moreover, 
the CPU wall-clock times simulated by SW and TW for the whole process with the resolution $dp=1/128$ are $144.10s$ and $89.14s$, respectively, and further $\alpha=0.64$, 
proving that the using the proposed kernel can significantly improve computational efficiency.
%
%
\subsection{Flow past a circular cylinder}
In the section, the flow around a circular cylinder is considered to test fluid-solid interaction with complex geometry.
To verify the simulation quantitatively, 
the drag and lift coefficient are given by
\begin{equation}\label{eq:wavespeed}
C_{D}=\frac{2F_{D}}{\rho_{\infty}u_{\infty}^2 A}\quad \text{and} \quad C_{L}=\frac{2F_{L}}{\rho_{\infty}u_{\infty}^2 A},
\end{equation}
where $F_{D}$ and $F_{L}$ are the drag and lift force respectively and $A$ is the size of the cross-sectional area of the cylinder.
In the case, 
the computational domain is $[25D, 15D]$ with the center of the cylinder located in $(7.5D, 7.5D)$ where the cylinder diameter $D = 2$, Strouhal number $St=fD/u_\infty$ and the Reynolds number $Re=\rho_{\infty}u_{\infty}D/\mu$ set as $100$. 
Besides, 
all boundaries are set as far-field boundary conditions and we apply three different resolutions $dp=1/10$, $1/20$ and $1/30$ to study its convergence.
\begin{figure}
    \centering
    \includegraphics[width=1.0\textwidth]{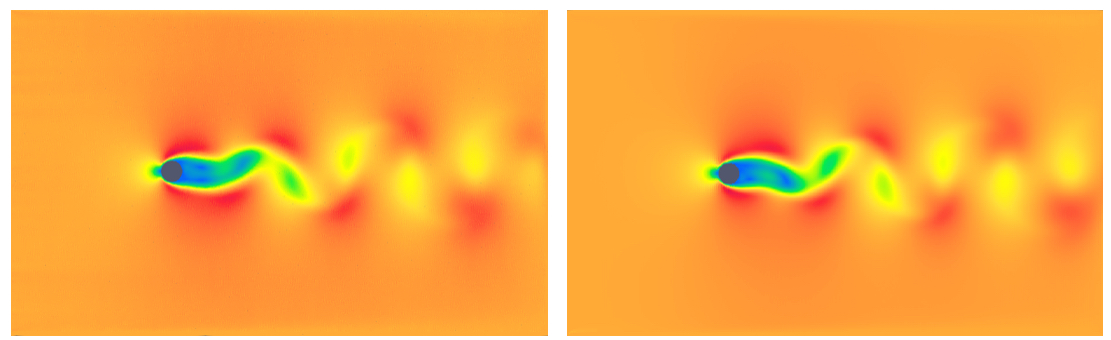}
    \caption{Flow around a cylinder: Velocity contour ranging from $1.8\times 10^{-3}$ to $1.5$ 
	using SW (left panel) and TW (right panel) kernels with $Re=100$ under the resolution $dp=1/10$ at $t=300$.}
    \label{flow around cylinder_velocity_contour}
\end{figure}
\begin{figure}
    \centering
    \includegraphics[width=1.0\textwidth]{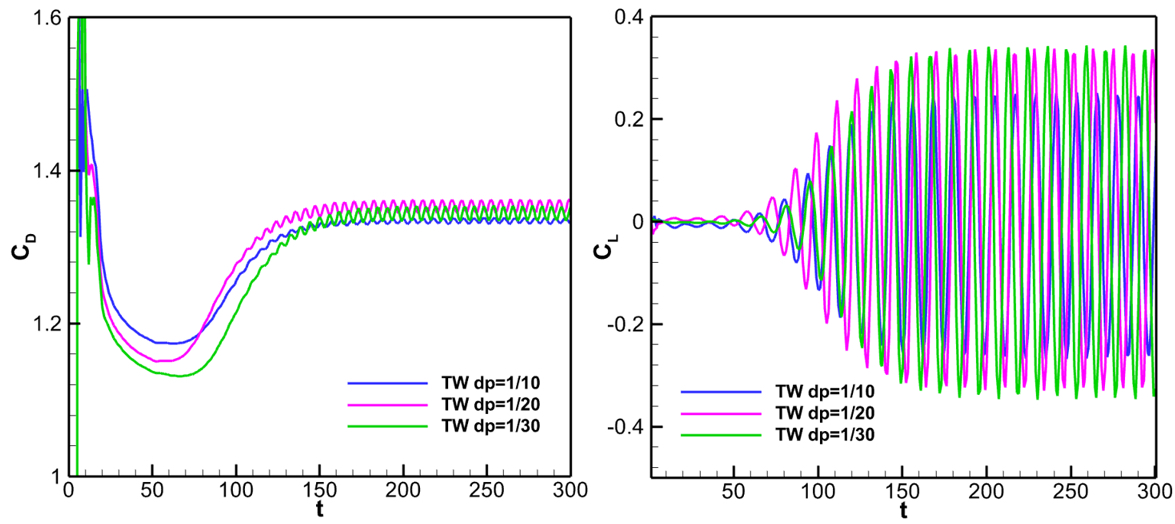}
    \caption{Flow around a cylinder: Drag (left panel) and lift (right panel) coefficients using truncated Wendland kernel with $Re=100$.}
    \label{flow around cylinder_coefficients}
\end{figure}
\begin{table}
    \caption{Flow around a cylinder: Drag and lift coefficients from different experimental and numerical results at Re=100. }
    \centering
    \begin{tabular}{cccc}
    \hline
    \begin{tabular}[c]{@{}c@{}}Parameters\end{tabular} &
    \begin{tabular}[c]{@{}c@{}}$C_{D}$\end{tabular} &
     \begin{tabular}[c]{@{}c@{}}$C_{L}$\end{tabular} &
     \begin{tabular}[c]{@{}c@{}}$S_{t}$\end{tabular} \\ \hline
    White\cite{white2006viscous} & 1.46 & -  & - \\ \hline
    Khademinejad et al.\cite{khademinezhad2015numerical} & 1.30 $\pm$ - & $\pm$0.292  & 0.152  \\ \hline
    Chiu et al.\cite{chiu2010differentially} & 1.35 $\pm$ 0.012 & $\pm$0.303  & 0.166  \\ \hline
    Le et al.\cite{le2006immersed} & 1.37 $\pm$ 0.009 & $\pm$0.323 & 0.160  \\ \hline
    Brehm et al.\cite{brehm2015locally} & 1.32 $\pm$ 0.010 & $\pm$0.320 & 0.165  \\ \hline
    Liu et al.\cite{liu1998preconditioned} & 1.35 $\pm$ 0.012 & $\pm$0.339 & 0.165  \\ \hline
    TW & 1.34 $\pm$ 0.010 & $\pm$0.340 & 0.172  \\ \hline
    \end{tabular}
    \label{Table_Re=100}
\end{table}
\begin{table}[]
\caption{Flow around a cylinder: Comparison of computational efficiency using SW and TW with different spatial resolutions.}
 \centering
\begin{tabular}{cccc}
\hline
Resolution     & Kernel type & CPU wall-clock time (s) & $\alpha$   \\ \hline
\multirow{2}{*}{1/10}  & SW      &    2672.98    & \multirow{2}{*}{0.73} \\
                       & TW      &    1949.30    &         \\
\multirow{2}{*}{1/20}  & SW      &    21397.50   & \multirow{2}{*}{0.72} \\
                       & TW      &    15412.23   &                      \\
\multirow{2}{*}{1/30} & SW      &     68340.33  & \multirow{2}{*}{0.72} \\
                       & TW      &    49256.36   &     \\ \hline
\end{tabular}
\label{Flow around a cylinder:computational efficiency.}
\end{table}

Figure \ref{flow around cylinder_coefficients} presents the drag and lift coefficients using truncated Wendland kernel with the resolutions $dp=1/10$, $1/20$ and $1/30$ under the Reynolds number $Re=100$, 
showing that the drag coefficients attain a stable value after an initial period of fluctuation, 
while the lift coefficient continues to oscillate around zero.
The converged result is documented in Table \ref{Table_Re=100}, alongside various other experimental and numerical findings \cite{white2006viscous,khademinezhad2015numerical,chiu2010differentially,le2006immersed,brehm2015locally,liu1998preconditioned} and the comparisons demonstrates the well agreement between the converged result obtained by proposed method and the referenced data, indicating its reasonability.
Furthermore, 
the CPU wall-clock times required by SW and TW in whole process with different resolutions are listed in Table \ref{Flow around a cylinder:computational efficiency.}, 
clearly showing that he computational time using truncated Wendland kernel is approximately 70\% of that using the standard Wendland kernel and the computational efficiency improves significantly by using the former.
%
%
\subsection{Twisting column}
In this section, the 3D simulation twisting column is further studied to verify the correctness of the proposed kernel. 
Following Refs. \cite{lee2016new,zhang2022artificial,wu2023essentially}, 
the column with the height $H=1m$ and length $L=6m$ is assumed to consist of a nearly incompressible material, characterized by a density of $\rho_0 = 1100 Kg/m^3$, Young's modulus and Poisson's ratio are $E=17 MPa$ and $\nu = 0.499$, respectively.
Besides, 
the initial condition for the twisting motion involves a sinusoidal rotational velocity field represented as $\mathbf{\omega}=[0, \Omega_{0}\sin{(\pi y_0/2L)}, 0] $ with $\omega_0 = 105 rad/s$.
\begin{figure}
    \centering
    \includegraphics[width=1.0\textwidth]{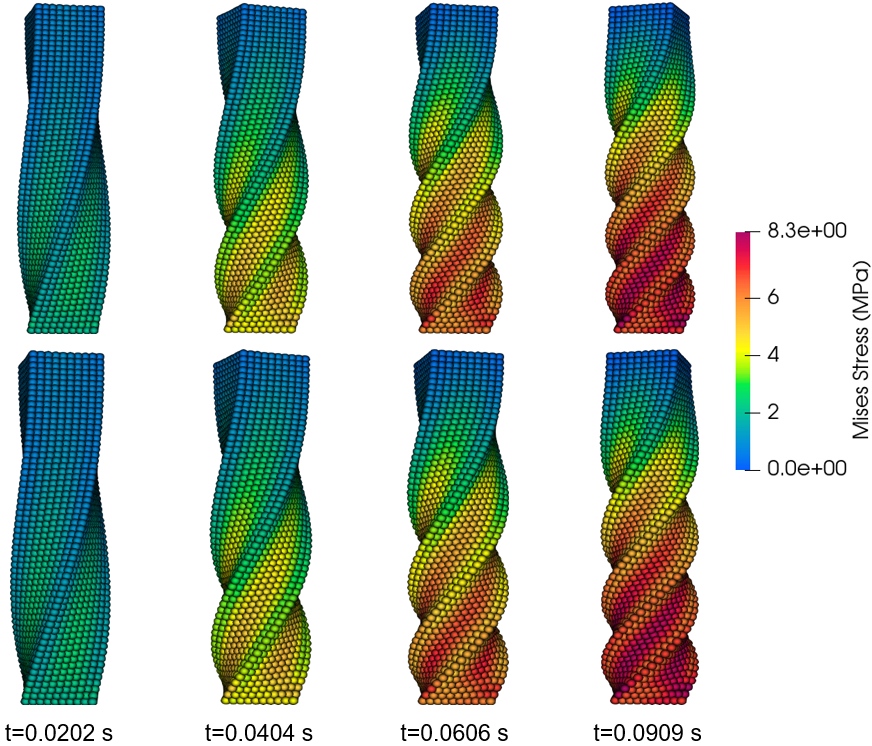}
    \caption{Twisting column: Mises stress ranging from $0MPa$ to $8.3MPa$ at different time instants obtained by SW (top panel) and TW (bottom panel) with initial rotational velocity $\mathbf{\omega}=[0, \Omega_{0}\sin{(\pi y_0/2L)}, 0] $ with $\omega_0 = 105$ rad/s under the spatial resolution $H/dp=10$ where $dp$ is the initial particle spacing.}
    \label{Twisting column Mises stress}
\end{figure}
\begin{figure}
    \centering
    \includegraphics[width=1.0\textwidth]{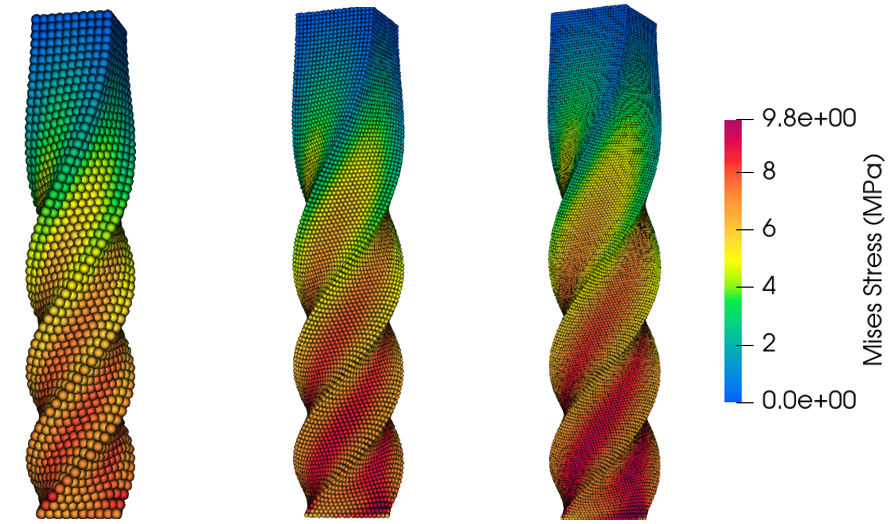}
    \caption{Twisting column: Mises stress ranging from $0MPa$ to $9.8MPa$ at different time instants obtained by TW with initial rotational velocity $\mathbf{\omega}=[0, \Omega_{0}\sin{(\pi y_0/2L)}, 0] $ with $\omega_0 = 105$ rad/s under the different spatial resolutions as $H/dp=10$ (left panel), $H/dp=20$ (middle panel) and $H/dp=30$ (right panel) where $dp$ is the initial particle spacing at the time $t=0.0909s$.}
    \label{Twisting column Mises stress convergence}
\end{figure}

Figure \ref{Twisting column Mises stress} portrays the Mises stress distribution of twisting column at different timings using SW and TW with $\omega_0 = 105$ rad/s under the resolution $H/dp=10$, 
showing that both kernels can obtain the smooth stress distribution which are agreement well with the reference \cite{wu2023essentially}.
Furthermore, 
three different resolutions are applied to verify its numerical stability shown in Figure \ref{Twisting column Mises stress convergence} and it can be observed that the smooth results can be obtained by TW with the increase of the resolutions and seem as stable.
To compare the computational efficiency between thses both kernels, 
the CPU wall-clock times required in the whole process are listed in Table \ref{Twisting column efficiency.}, 
indicating that the considerably improved computational efficiency can be achieved by the proposed kernel.

\begin{table}[]
\caption{Twisting column: Comparison of computational efficiency using SW and TW with different spatial resolutions.}
 \centering
\begin{tabular}{cccc}
\hline
Resolution     & Kernel type & CPU wall-clock time (s) & $\alpha$   \\ \hline
\multirow{2}{*}{1/10}  & SW      &    27.55  & \multirow{2}{*}{0.67} \\
                       & TW      &    18.52   &     \\
\multirow{2}{*}{1/20}  & SW      &    366.91  & \multirow{2}{*}{0.68} \\
                       & TW      &    247.94  &                      \\
\multirow{2}{*}{1/30} & SW      &     1825.15  & \multirow{2}{*}{0.64} \\
                       & TW      &    1173.56   &     \\ \hline
\end{tabular}
\label{Twisting column efficiency.}
\end{table}
%
%
%
\subsection{Bending column}
In this section, 
a $3D$ bending column is considered to test the accuracy and computational efficiency of the proposed method.
Following Refs. \cite{zhang2021integrative,wu2023essentially}, 
in this case, we employ a rubber-like material with a length of $L=6m$ and a height of $H = 1m$ and the material's cross-sectional area is square in shape and remains fixed at the bottom during the simulation.
Besides, 
the initial velocity is $\mathbf{v}_{0}=10(\frac{\sqrt 3}{2},\frac{1}{2},0)^{T}$, density $\rho_{0}=1100kg/m^{3}$, 
Young's modulus $E=17MPa$ and Poisson's ratio $\nu = 0.45$.
For the convergence study, 
we apply three different resolutions $dp=1/6$, $1/12$ and $1/24$.
\begin{figure}
    \centering
    \includegraphics[width=1.0\textwidth]{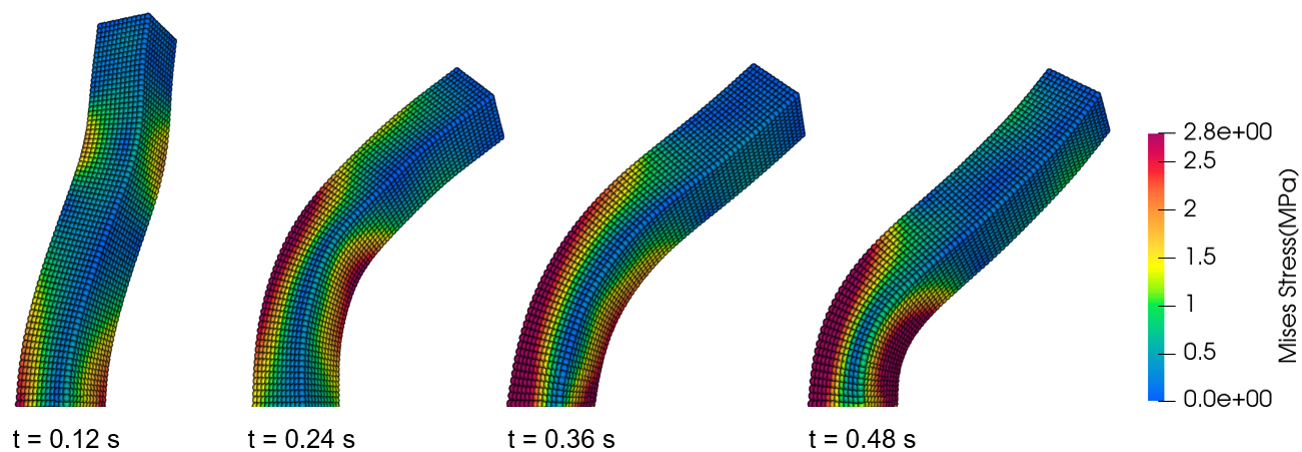}
    \caption{Bending column: The Mises stress from $0MPa$ to $2.8MPa$ using TW with the spatial resolution $H/dp=12$ at the different instants.}
    \label{Bending column Miese stress}
\end{figure}
\begin{figure}
    \centering
    \includegraphics[width=1.0\textwidth]{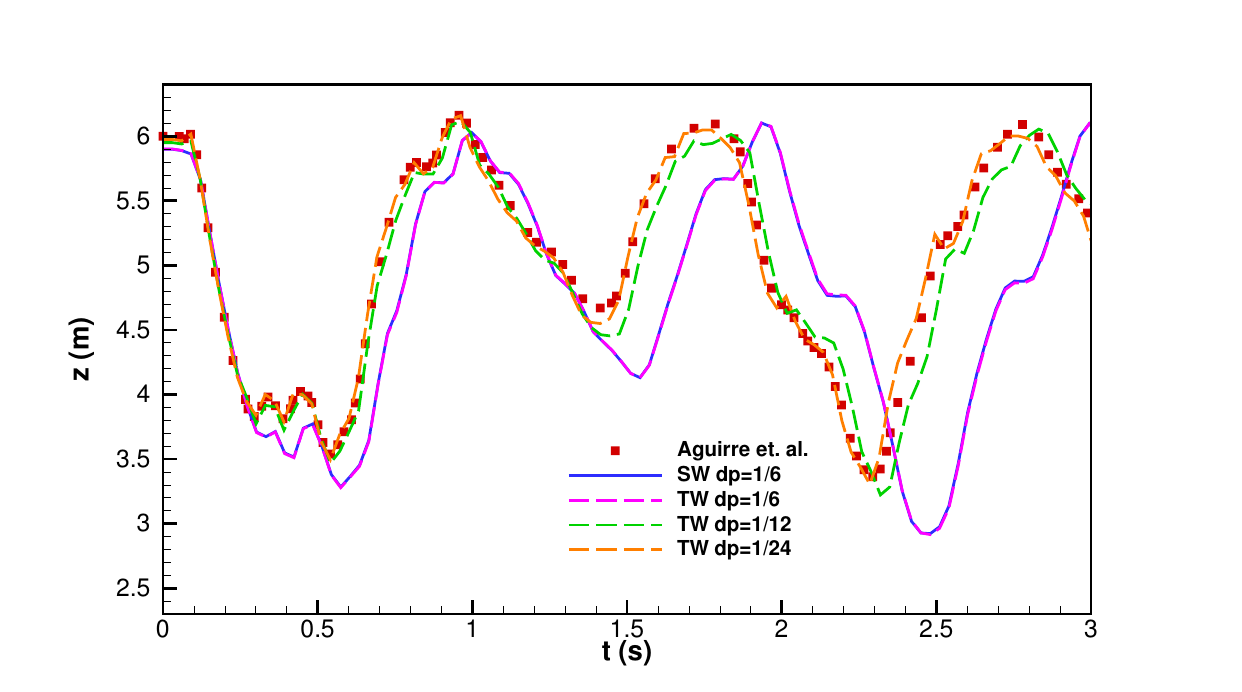}
    \caption{Bending column: The vertical position z located at the top of the column \cite{wu2023essentially} 
	using SW and TW with three different resolutions 
	and the comparisons with the reference data from Aguirre et al. \cite{aguirre2014vertex}.}
    \label{3D Bending column}
\end{figure}
\begin{table}[]
\caption{Bending column: Comparison of computational efficiency using SW and TW with different spatial resolutions.}
 \centering
\begin{tabular}{cccc}
\hline
Resolution     & Kernel type & CPU wall-clock time (s) & $\alpha$   \\ \hline
\multirow{2}{*}{1/6}  & SW      &    1.45 & \multirow{2}{*}{0.69} \\
                       & TW     &    1.01  &         \\
\multirow{2}{*}{1/12}  & SW     &    27.64 & \multirow{2}{*}{0.47} \\
                       & TW     &    13.12 &                      \\
\multirow{2}{*}{1/24} & SW      &    626.31 & \multirow{2}{*}{0.56} \\
                       & TW     &    352.11 &     \\ \hline
\end{tabular}
\label{3D Bending column:computational efficiency.}
\end{table}

Figure \ref{Bending column Miese stress} presents the Mises stress from $0MPa$ to $2.8MPa$ using TW with the spatial resolution $H/dp=12$ at the different timings, 
clearly showing that the smooth results can be obtained by the proposed kernel well.
Besides, 
Figure \ref{3D Bending column} depicts the vertical position $z$ located at the top of the column \cite{wu2023essentially} using SW and TW with three different resolutions and the comparisons with the reference data from Aguirre et al. \cite{aguirre2014vertex}.
From the results, 
the results using two methods at the resolution $dp=1/6$ are almost coincide, 
verifying that these methods enable obtain the same accuracy.
Also, 
with the increase of the spatial resolutions, 
the results converge rapidly and achieve second order convergence.
More importantly, 
the computational efficiency using SW and TW are listed in Table \ref{3D Bending column:computational efficiency.}, 
showing that the computational time using the latter is almost half of that using the former, i.e. significantly improved computational efficiency using truncated Wendland kernel.
%
%
\section{Summary and conclusion}\label{Summary and conclusions}
In this paper, 
the truncation error of the Wendland kernel, 
with its compact support truncated to $1.6h$, 
has been subject to comprehensive quantitative analysis, 
and the results have demonstrated that the error when using the truncated Wendland kernel is comparable to that of the standard Wendland kernel 
in conjunction with the kernel gradient correction.
A series of numerical tests encompassing both fluid and solid dynamics have been conducted, 
and clearly show that the proposed truncation scheme results in significantly improved computational efficiency without compromising numerical accuracy. 
Note that the extent of this improvement does vary depending on the specific scenario.
Overall, 
the computational time required for calculations involving the truncated Wendland kernel has been reduced by a minimum of 30\% compared to the demands of the standard Wendland kernel.
%
%
\bibliographystyle{elsarticle-num}
\bibliography{ref}
\end{document}